\documentclass[showpacs,12pt,a4paper]{article}
\usepackage{amsmath}
\usepackage{amsmath,amssymb,epsfig,fleqn,url,psboxit,color,here,graphics,graphicx,rotating,multirow}

\begin{document}
\textwidth=135mm
 \textheight=200mm
\begin{center}
{\bfseries Six "beyond Collins and Sivers" transverse spin asymmetries at COMPASS
\footnote{{\small Contribution given at the 20th International Symposium on Spin Physics (SPIN2012)
JINR, Dubna, Russia, September 17 - 22, 2012}}}
\vskip 5mm
B. Parsamyan$^{\dag,\ddag}$
\vskip 5mm
{\small {\it $^\dag$ Dipartimento di Fisica Generale, Universit\`a di Torino, Torino, Italy}} \\
{\small {\it $^\ddag$ INFN, Sezione di Torino, Via P. Giuria 1, I-10125 Torino, Italy}}
\\
\end{center}
\vskip 5mm
\centerline{\bf Abstract}
COMPASS is a fixed-target high energy physics experiment at the SPS
at CERN \cite{Abbon:2007pq}. One of the important objectives of the experiment is the exploration of the transverse spin structure of the nucleon via spin dependent azimuthal asymmetries in single-hadron production in deep inelastic scattering of polarized leptons off transversely polarized target. For this purpose a series of measurements were made in COMPASS, using 160 GeV/c longitudinally polarized muon beam and transversely polarized $^6LiD$ (in 2002, 2003 and 2004) and $NH_3$ (in 2007 and 2010) targets.

Till now main attention was focused on Collins and Sivers asymmetries and obtained results play an important role in the general understanding of the three-dimensional nature of the nucleon in terms of Transverse Momentum Dependent (TMD) Parton Distribution Functions (PDFs) and Fragmentation Functions (FFs).

In addition to these two measured leading-twist effects, the SIDIS cross-section counts six more target transverse spin dependent azimuthal effects, which have their own well defined leading or higher-twist interpretation in terms of QCD parton model. So far COMPASS presented preliminary results for these asymmetries from deuteron \cite{Parsamyan:2007ju},\cite{Kotzinian:2007uv} and "proton-2007" data \cite{Parsamyan:2010se}. In this contribution we review the results obtained with the last "proton-2010" data sample.
\\

\vskip 10mm
\section{\label{sec:intro}Introduction}

\begin{figure}[h]
\center
\includegraphics[width=1.0\textwidth]{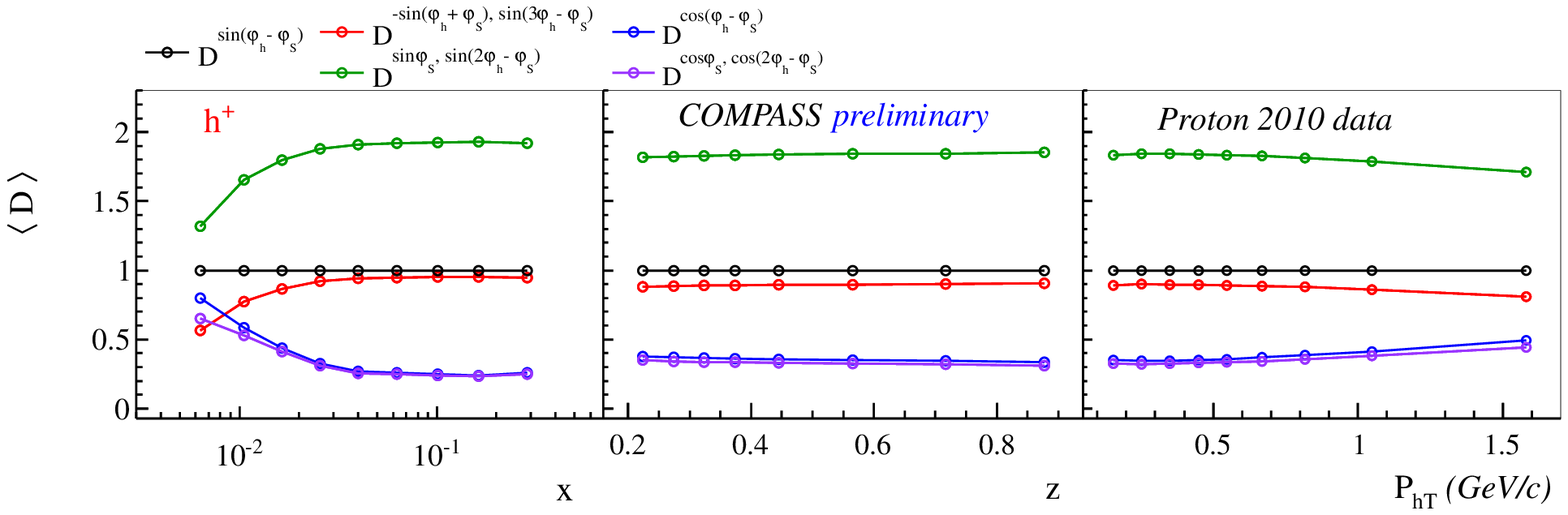}
\includegraphics[width=1.0\textwidth]{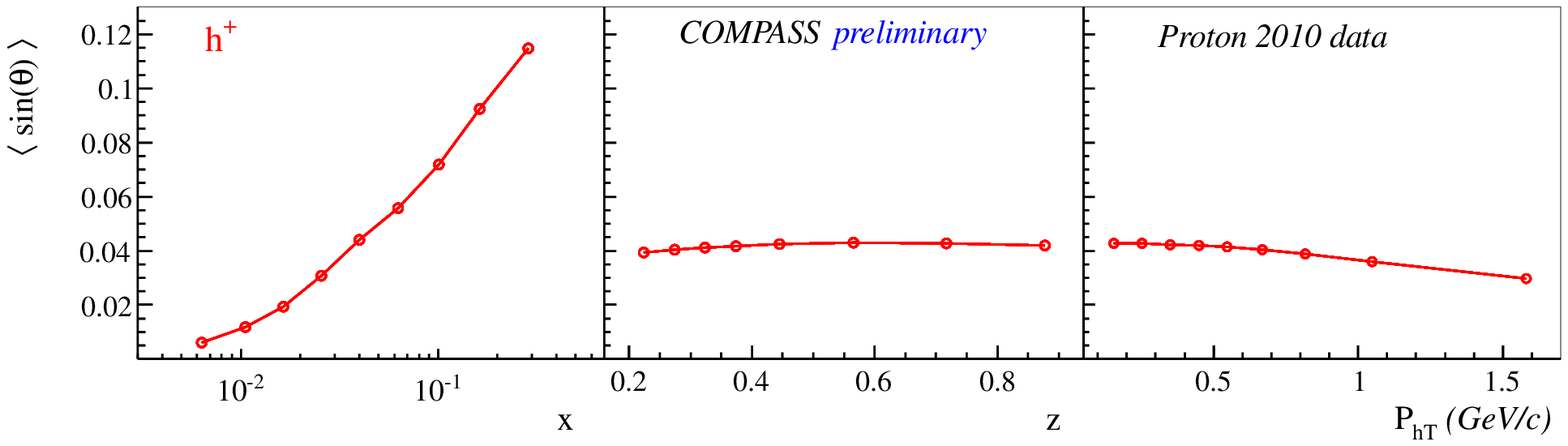}
\includegraphics[width=1.0\textwidth]{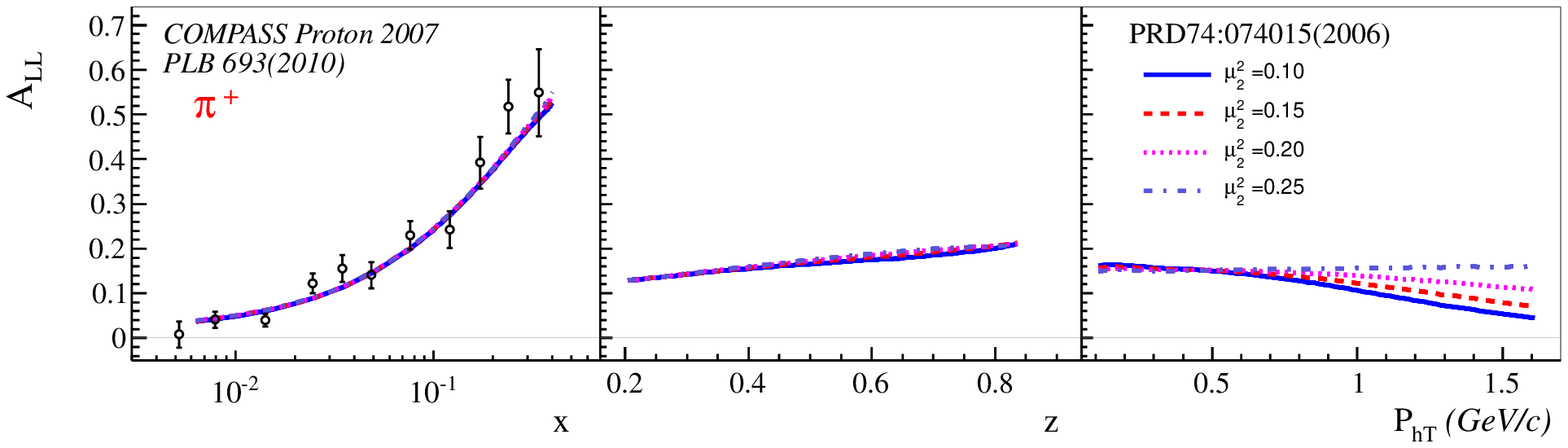}
\caption{Mean $D$ (depolarization) and $sin(\theta)$ factors and $A_{LL}$ asymmetry.}
\label{fig:DySinTHA_LL}
\end{figure}

Following the standard SIDIS definitions from \cite{Kotzinian:1994dv},\cite{Bacchetta:2006tn} and taking into account the corrections due to the difference between target transverse polarization defined relative to the lepton beam ($P_T$) or to the virtual photon direction ($S_T$) \cite{Diehl:2005pc}, the cross-section expression for transversely (in lab. system) polarized target can be re-written in the following way \cite{Kotzinian:1994dv}-\cite{Diehl:2005pc} \footnote{The notations are equivalent to those used in \cite{Bacchetta:2006tn}, \cite{Parsamyan:2007ju}-\cite{Parsamyan:2010se} and $\theta$ is the angle between $\gamma^*$-direction and initial lepton momenta}:
{\footnotesize
\begin{eqnarray}
\label{eq:x_sec_mod}
  && \hspace*{-1.5cm}\frac{{d\sigma }}{{dxdydzdP_{hT}^2d{\varphi _h}d{\varphi _S}}} = \\
 &&\hspace*{0.1cm}  \left[ {\frac{{\cos \theta }}{{1 - {{\sin }^2}\theta {{\sin }^2}{\varphi _S}}}} \right]  \left[ {\frac{\alpha }{{xy{Q^2}}}\frac{{{y^2}}}{{2\left( {1 - \varepsilon } \right)}}\left( {1 + \frac{{{\gamma ^2}}}{{2x}}} \right)} \right]  \left( {{F_{UU,T}} + \varepsilon {F_{UU,L}}} \right) \times \nonumber
\end{eqnarray}
   \[\begin{gathered}[h!]\hspace*{-0.85cm} \left( \begin{gathered}
  1 + \cos {\varphi _h}  \sqrt {2\varepsilon \left( {1 + \varepsilon } \right)} A_{UU}^{\cos {\varphi _h}} + \cos {2{\varphi _h}}  \varepsilon  A_{UU}^{\cos {2{\varphi _h}}} +  \hfill \\
  \hspace*{4.5cm} \lambda \sin {\varphi _h}  \sqrt {2\varepsilon \left( {1 - \varepsilon } \right)} A_{LU}^{\sin {\varphi _h}} +
  \left[{{\sqrt {1 - {{\sin }^2}\theta {{\sin }^2}{\varphi _S}} }} \right]^{-1}\times\hfill\\
  {{\text{P}}_{\text{T}}} \left[ \begin{gathered}
  \sin {\varphi _S}  \left( \cos \theta {\sqrt {2\varepsilon \left( {1 + \varepsilon } \right)}  A_{UT}^{\sin {\varphi _S}}} \right) +  \hfill \\
  \sin \left( {{\varphi _h} - {\varphi _S}} \right)  \left( {\cos \theta A_{UT}^{\sin \left( {{\varphi _h} - {\varphi _S}} \right)} + \frac{1}{2}\sin \theta\sqrt {2\varepsilon \left( {1 + \varepsilon } \right)} A_{UL}^{\sin {\varphi _h}}} \right) +  \hfill \\
  \sin \left( {{\varphi _h} + {\varphi _S}} \right)  \left(\cos \theta  {\varepsilon A_{UT}^{\sin \left( {{\varphi _h} + {\varphi _S}} \right)} + \frac{1}{2}\sin \theta \sqrt {2\varepsilon \left( {1 + \varepsilon } \right)} A_{UL}^{\sin {\varphi _h}}} \right) +  \hfill \\
  \sin \left( {2{\varphi _h} - {\varphi _S}} \right)  \left( \cos \theta{\sqrt {2\varepsilon \left( {1 + \varepsilon } \right)}  A_{UT}^{\sin \left( {2{\varphi _h} - {\varphi _S}} \right)} + \frac{1}{2}\sin \theta \varepsilon A_{UL}^{\sin 2{\varphi _h}}} \right) +  \hfill \\
  \sin \left( {3{\varphi _h} - {\varphi _S}} \right)  \left(\cos \theta {\varepsilon  A_{UT}^{\sin \left( {3{\varphi _h} - {\varphi _S}} \right)}} \right) +
    \sin \left( {2{\varphi _h} + {\varphi _S}} \right)  \left( {\frac{1}{2}\sin \theta \varepsilon A_{UL}^{\sin 2{\varphi _h}}} \right) \hfill \\
\end{gathered}  \right] +  \hfill \\
  {{\text{P}}_{\text{T}}}{\lambda} \left[ \begin{gathered}
  \cos {\varphi _S}  \left( \cos \theta {\sqrt {2\varepsilon \left( {1 - \varepsilon } \right)} A_{LT}^{\cos {\varphi _S}} +  \sin \theta \sqrt {\left( {1 - {\varepsilon ^2}} \right)} {A_{LL}}} \right) +  \hfill \\
  \cos \left( {{\varphi _h} - {\varphi _S}} \right)  \left( \cos \theta {\sqrt {\left( {1 - {\varepsilon ^2}} \right)} A_{LT}^{\cos \left( {{\varphi _h} - {\varphi _S}} \right)} + \frac{1}{2} \sin \theta \sqrt {2\varepsilon \left( {1 - \varepsilon } \right)} A_{LL}^{\cos {\varphi _h}}} \right) +  \hfill \\
    \cos \left( {2{\varphi _h} - {\varphi _S}} \right)  \left( \cos \theta {\sqrt {2\varepsilon \left( {1 - \varepsilon } \right)} A_{LT}^{\cos \left( {2{\varphi _h} - {\varphi _S}} \right)}} \right) + \hfill \\
  \cos \left( {{\varphi _h} + {\varphi _S}} \right)  \left( {\frac{1}{2}\sin \theta \sqrt {2\varepsilon \left( {1 - \varepsilon } \right)} A_{LL}^{\cos {\varphi _h}}} \right) \hfill
\end{gathered}  \right] \hfill
\end{gathered}  \right) \hfill
\end{gathered} \]
}
There are new $sin\theta$-scaled terms and $\theta$-depending factors and two new modulations ($\sin \left( {2{\varphi _h} + {\varphi _S}} \right)$ and $\cos \left( {{\varphi _h} + {\varphi _S}} \right)$) appearing in this cross-section expression compared with the one presented, for instance, in \cite{Parsamyan:2007ju}-\cite{Parsamyan:2010se}, in which the effects due to the $P_T$ to $S_T$ transition have been neglected.
The equation (\ref{eq:x_sec_mod}) counts in total eight: five Single-Spin (SSA) and three Double-Spin (DSA) target transverse polarization dependent asymmetries. Since the $sin\theta$ is rather small quantity in COMPASS kinematics (see Fig.\ref{fig:DySinTHA_LL}) influence of the additional terms and factors can be neglected in case of all the asymmetries except for $A_{LT}^{\cos {\varphi _S}}$ DSA, which, even taking into account suppression by a $sin\theta$ scale-factor, is still sizably affected by large $A_{LL}$ amplitude \cite{Alekseev:2010ub}.
In Fig.\ref{fig:DySinTHA_LL} the theoretical curves for $A_{LL}$, evaluated based on \cite{Anselmino:2006yc} and used for the correction of $A_{LT}^{\cos {\varphi _S}}$ asymmetry, are compared with the COMPASS data points \cite{Alekseev:2010ub}, demonstrating close agreement.

The eight target transverse spin dependent "raw" asymmetries are extracted simultaneously, using unbinned maximum likelihood technique and then are corrected for the $D$ depolarization factors ($\varepsilon$-depending factors in equation (\ref{eq:x_sec_mod}) standing in front of the amplitudes), dilution factor and target and beam (only DSAs) polarizations \cite{Parsamyan:2007ju},\cite{Parsamyan:2010se}. Measured in COMPASS mean $D$ factors corresponding to different asymmetries are presented in Fig.\ref{fig:DySinTHA_LL}.

In the QCD parton model approach four of the eight transverse spin asymmetries ($A_{UT}^{sin(\phi_h+\phi_S)}$, $A_{UT}^{sin(\phi_h+\phi_S)}$, $A_{UT}^{\sin (3\phi _h -\phi _s )}$ SSAs and $A_{LT}^{\cos (\phi _h -\phi _s )}$ DSA) have Leading Order(twist) (LO) interpretation and are described by the convolutions of twist-two TMD PDFs and FFs, while the other four ($A_{UT}^{\sin (\phi _s )}$ and $A_{UT}^{\sin (2\phi _h -\phi _s )}$ SSAs and $A_{LT}^{\cos (\phi _s)}$ and $A_{LT}^{\cos (2\phi _h -\phi _s )}$ DSAs), despite their higher-twist origin, however, can be represented as "Cahn kinematic corrections" to twist-two effects.
These sub-leading amplitudes are suppressed with respect to the leading twist ones by $\sim M/Q$ (for details see: \cite{Bacchetta:2006tn},\cite{Mulders:1995dh},\cite{Parsamyan:2007ju}-\cite{Parsamyan:2010se}).
It can be shown that LO $A_{UT}^{\sin (3\phi _h -\phi _s )}$ (related to the $h^{q\perp}_{1T}$ "pretzelosity" PDF) is expected to scale according to $\sim|\bf{P_{hT}}^3|$ and thus is suppressed by $\sim|\bf{P_{hT}}|^2$ w.r.t $\sim|\bf{P_{hT}}|$-scaled Collins, Sivers and $A_{LT}^{\cos (\phi _h -\phi _s )}$ LO amplitudes. Similarly, other four asymmetries are suppressed by $\sim|\bf{P_{hT}}|$.
The $A_{LT}^{\cos (\phi _h -\phi _s )}$ amplitude (related to the $g^{q\perp}_{1T}$ "worm gear" PDF) is of particular interest because it is the only transverse DSA expected to be sizable (LO, no suppression).

For Collins and Sivers effects, in addition to the previous measurements with deuteron and proton, COMPASS has recently published results from 2010 proton data \cite{Adolph:2012sn},\cite{Adolph:2012sp}. In the next section we present the preliminary results for the other six asymmetries obtained with the same data sample.
\section{Data analysis and results}
\begin{figure}[H]
\center
\includegraphics[width=0.975\textwidth]{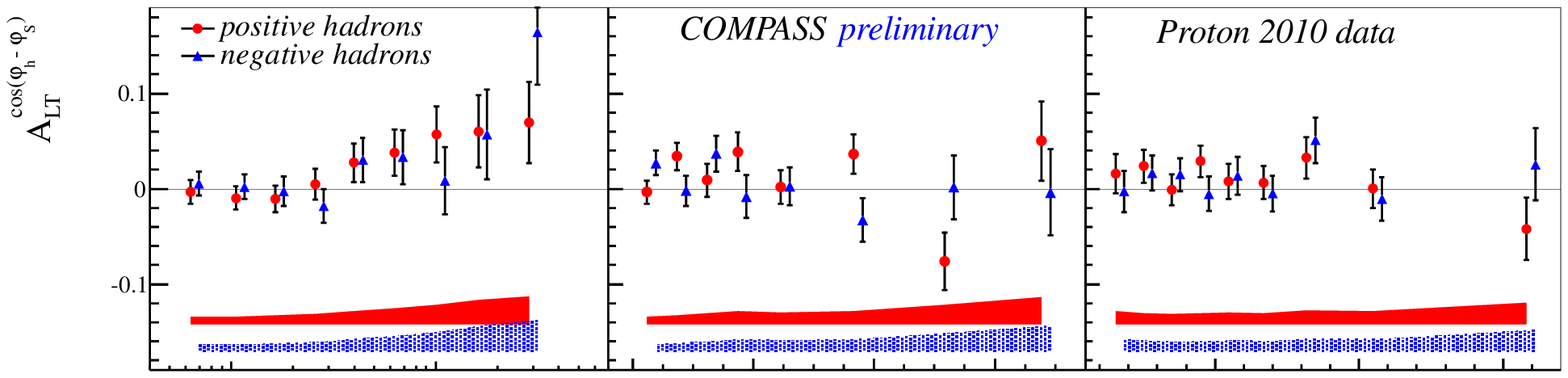}
\includegraphics[width=0.975\textwidth]{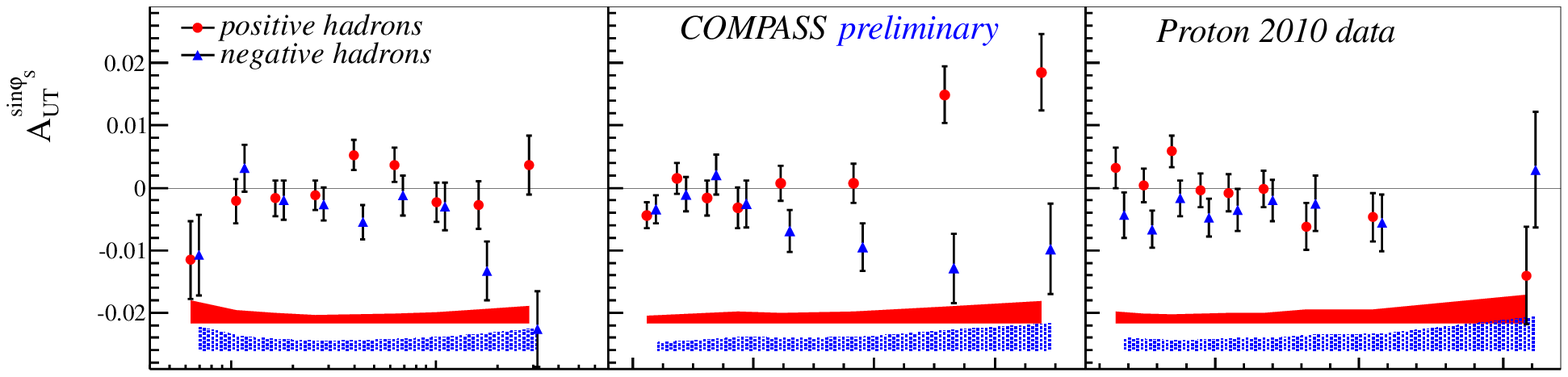}
\includegraphics[width=0.975\textwidth]{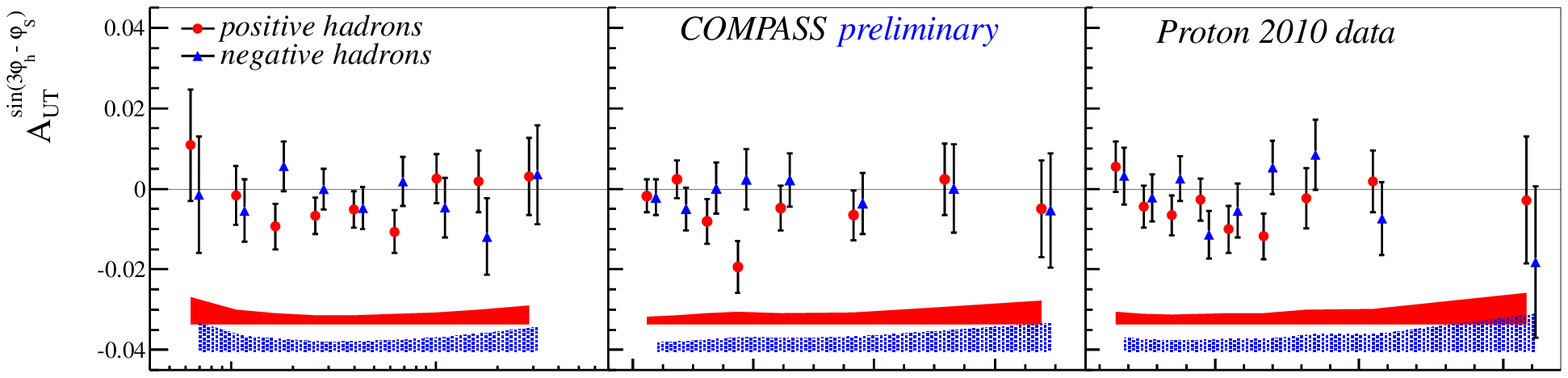}
\includegraphics[width=0.975\textwidth]{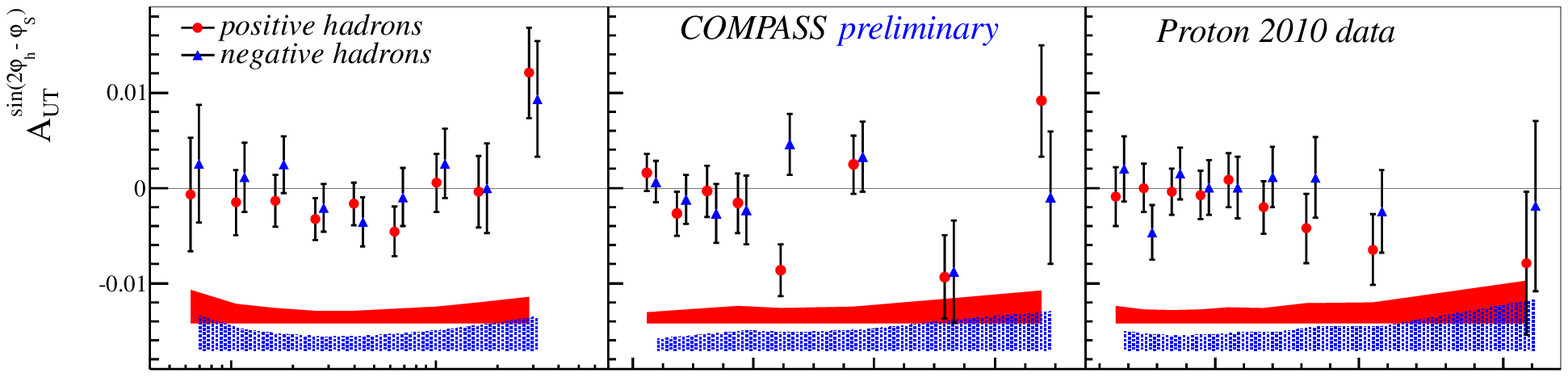}
\includegraphics[width=0.975\textwidth]{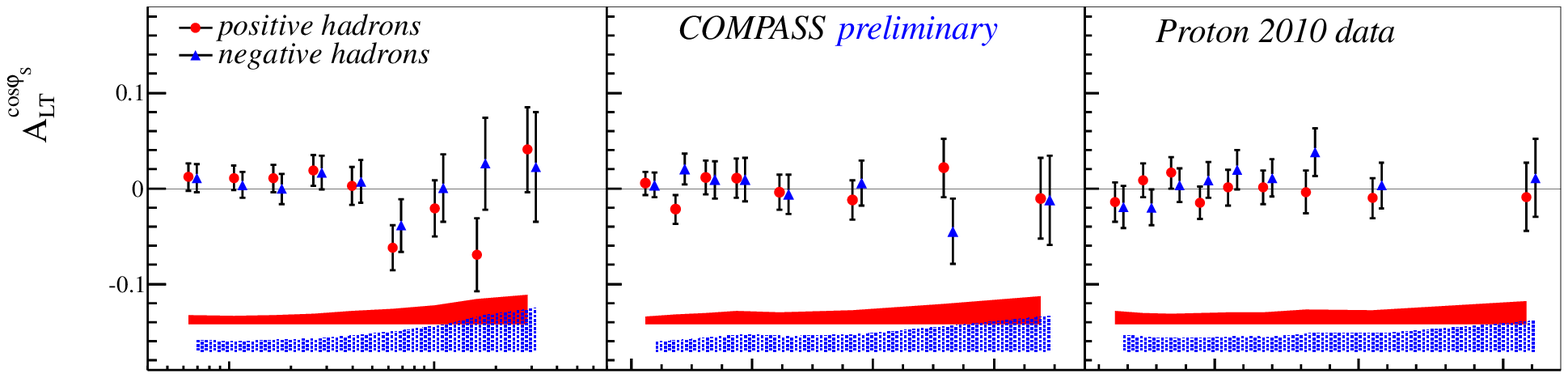}
\includegraphics[width=0.975\textwidth]{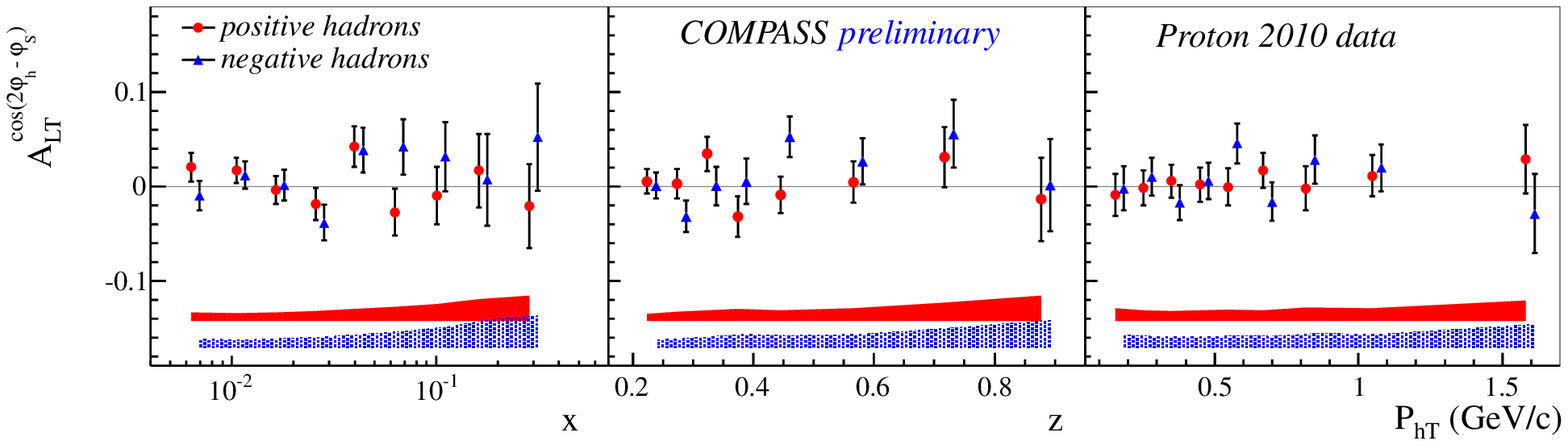}
\caption{Six "Beyond Collins and Sivers" asymmetries at COMPASS.}
\label{fig:A8}
\end{figure}

\begin{figure}[h]
\center
\includegraphics[width=0.975\textwidth]{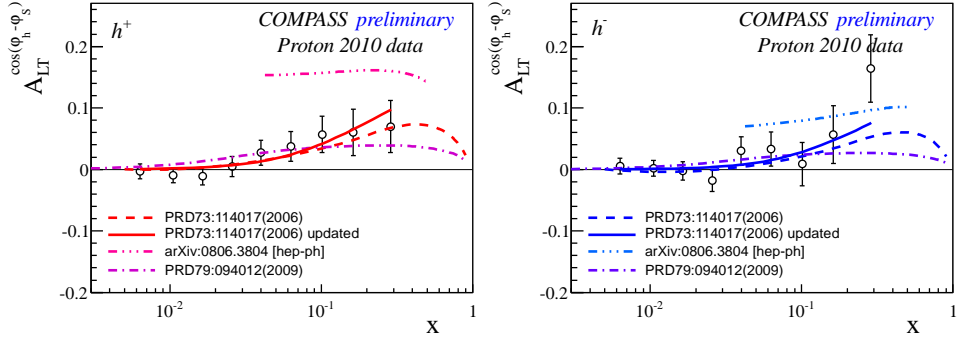}
\caption{$A_{LT}^{cos(\phi_h-\phi_S)}$ asymmetry: comparison with theories.}
\label{fig:A_LT}
\end{figure}

The whole data selection and analysis procedure applied for the extraction of six mentioned asymmetries from COMPASS 2010 proton data
is identical to the one applied in case of already published Collins and Sivers asymmetries.
The detailed description of COMPASS spectrometer and details on analysis can be found in: \cite{Abbon:2007pq},\cite{Parsamyan:2007ju},\cite{Parsamyan:2010se},\cite{Adolph:2012sn},\cite{Adolph:2012sp} (and references therein).
%
%
The asymmetries extracted as functions of $x$,$z$ and $P_{hT}$ for positive and negative hadron production are presented in Fig.\ref{fig:A8}. The systematic uncertainties for each asymmetry have been estimated separately for positive and negative hadrons and are given by the bands.
According to preliminary observations, there is an evidence of non-zero LO $A_{LT}^{\cos (\phi _h -\phi _s )}$ DSA and sub-leading $A_{UT}^{\sin (\phi _s )}$ SSA , while the other four "beyond Collins and Sivers" amplitudes are found to be compatible with zero within the statistical accuracy. It has to be mentioned that similar behavior for both non-zero amplitudes (and no effect for others) has been preliminary reported also by the HERMES collaboration.
In Fig.\ref{fig:A_LT} $A_{LT}^{\cos (\phi _h -\phi _s )}$ asymmetry, extracted from COMPASS 2010 proton data, is compared with the theoretical predictions from \cite{Kotzinian:2006dw}, \cite{Boffi:2009sh} and \cite{Kotzinian:2008fe}, demonstrating a good level of agreement between theory and measurement within the given statistical accuracy.
All the obtained results will be the subject of a future publication.
\section{Conclusions}\label{concl}
The preliminary results on six, additional to Collins and Sivers amplitudes, asymmetries from COMPASS proton 2010 data, have been presented. A non-zero trend has been observed for the $A_{LT}^{\cos (\phi _h -\phi _s )}$ and $A_{UT}^{\sin (\phi _s )}$ amplitudes, while the other four are found to be consistent with zero within the statistical accuracy. The measured kinematical dependencies of $A_{LT}^{\cos (\phi _h -\phi _s )}$ asymmetry are inline with the predictions given by several theoretical models.
Combined with the previous COMPASS measurements and data from other experiments, these results give another possibility to access TMD PDFs and FFs, and to study the spin-structure of the nucleon.

\end{document}